\documentclass[preprint]
{revtex4-2}

\usepackage[english]{babel}

\makeatletter
\addto\captionsenglish{\def\l@en{\l@english}}
\makeatother
\usepackage{graphicx}
\usepackage{dcolumn}
\usepackage{bm}
\usepackage{gensymb}
\usepackage{amsmath}
\usepackage{mathtools}
\usepackage{soul}
\usepackage{lineno}
\usepackage{natbib}
\usepackage{amsfonts}
\usepackage{amssymb}
\usepackage{mathrsfs}
\usepackage{bbold}
\usepackage{latexsym}
\usepackage[table,xcdraw]{xcolor}
\usepackage[normalem]{ulem}
\usepackage{braket}
\usepackage{subfigure}
\usepackage{float}
\usepackage{comment}
\usepackage{physics}
\usepackage{hyperref}
\usepackage{lineno}
\usepackage{physics}
\usepackage{lineno}
\usepackage{amssymb}
\usepackage{float}
\usepackage{multirow}

\newcommand{\Hcal}{\mathcal{H}}
\newcommand{\Dcal}{\mathcal{D}}

\newcommand{\Ecal}{\mathcal{E}}

\begin{document}
	
	
	\title{ Simulation of depolarizing channel exploring maximally non separable spin-orbit mode }
	
	

	\author{G.Tiago$^{1,2}$, V.S. Lamego $^{3}$, M.H.M Passos$^{4}$, W. F. Balthazar$^{1,5}$, }%
 \author{ J. A. O. Huguenin$^{1,2}$}
	\email{jose$_$hugguenin@id.uff.br}
	
	\affiliation{%
		$^1$ Programa de Pós-graduação em Física - Instituto de F\'{\i}sica - Universidade Federal Fluminense - Niter\'{o}i -- RJ -- 24210-346, Brazil\\
		$^2$	Instituto de Ciencias Exatas - Universidade Federal Fluminense - Volta Redonda -- RJ -- 27213-145 Brazil \\
      $^3$  Instituto Federal de Educação, Ciência e Tecnologia do Paraná, Assis Chateaubrian -- PR -- 85935-000, Brazil \\
      Volta Redonda -- RJ -- 27213-100 Brazil \\
      $^4$ Centro Brasileiro de Pesquisas Físicas, Rua Dr. Xavier Sigaud, 150, Urca, Rio de Janeiro, 22290-180, RJ, Brazil\\
		$^5$	Instituto Federal de Educação, Ciência e Tecnologia do Rio de Janeiro - IFRJ --Volta Redonda -- RJ 27213-100, Brazil\\
	}%
	
	
	
	
	\date{\today}
	
\begin{abstract}	
The Depolarizing channel is one of the most important noise models and constitutes a reliable benchmark for the quantum information field. In this work, we present a simple way to emulate a Depolarizing channel by exploring a maximally non separable spin-orbit mode in a compact linear optical circuit. The evolution of different states has been successfully reproduced. Our results are in excellent agreement compared with the results obtained by the spin-orbit Solovay-Kitaev decomposition for the Depolarizing channel, also presented here for the first time. Our proposal can be a powerful tool for  studies of depolarizing effect. %
\end{abstract}
	
	\maketitle
	
	
%
\section{Introduction}
Quantum information processes are susceptible to decoherence, the main challenge for the realization of quantum information and quantum computation protocols  \cite{schlosshauer2019quantum}. Modeling decoherence processes in quantum channels has pivotal relevance once quantum networks require a reliable channel for entanglement or quantum superposition distribution \cite{bruss2000quantum, riccardi2021simultaneous}.  Furthermore, it is very important for the study of the classical-quantum transition\cite{zurek2003decoherence}.

Even Depolarizing Channel presents deleterious action on the standard model of quantum Shannon theory, a quantum communication network was achieved by combining two depolarizing channels to transmit classical information when they are combined in a quantum superposition \cite{ebler2018enhanced}. Such as the prediction of transmitting information by means noise channel was verified experimentally \cite{goswami2020increasing, guo2020experimental}. Perfect quantum communication can be envisaged by using entanglement-breaking channels \cite{chiribella2021indefinite}. Also, quantum information can be transmitted by using N completely depolarizing channels in a superposition of alternative causal orders \cite{chiribella2021quantum}. Such results reinforce the relevance of studies on depolarizing channels.

On the other hand, the degrees of freedom (DoF) of light have proven to be a very robust platform to explore quantum features. By using polarization and transverse mode DoF, we can build the well-known spin-orbit modes \cite {topo} that present the same mathematical structure as a two-system entangled state \cite{Pereira}. These Bell-like modes violate Bell's inequalities written for quantum entangled states \cite{bell1, bell3} and can be seen as a suitable test bed apparatus for quantum computation through proof of principle for quantum gates \cite{cqg1, cqg3}. In quantum communication, a BB84 protocol for quantum cryptography without a shared reference frame was experimentally demonstrated \cite{ccrypt}. In the investigation of the fundamentals of quantum mechanics, we can cite system-environment interaction \cite{environ}, contextuality \cite{li2017experimental, context}, quantum thermodynamics \cite{quantThermoOPTSIM}, and transition from quantum-to-classical random walk \cite{lamego2024transition}. By adding path degree of freedom, a tripartite entangled state can be emulated \cite{tript}. Mixed states are also emulated by spin-orbit modes, as X-states \cite{Xstate, MaxiDiscordantSOM} that were used to experimentally measure Discord on an intense laser beam \cite{DiscordSpinOrbitModes}.

Degrees of freedom of light of intense laser beams were used to emulate a quantum channel both in the Markovian \cite{obando2020simulating} and Non-Markovian \cite{passos2019non} regime. A very complete simulation of a quantum channel with an intense laser beam was the implementation of the Solovay-Kitaev (SK) decomposition by using transverse mode as an ancillary system \cite{SolovayKitaevPassos}. However, the depolarizing channel was not emulated. 

In this paper, we present a very compact linear optical circuit to emulate the dynamics of a single qubit in a depolarizing channel.  We compare the results with the implementation of SK-decomposition, with excellent agreement showing more robustness in the results. The paper is organized as follows. Section II presents a theoretical background. Section III presents the experimental proposal. The results are presented in Section IV, where we also compare with unprecedented results of the Depolarizing Channel with SK-decomposition. Concluding remarks are presented in Section V.

 \section{Theory}

\subsection{Depolarizing Channel}

The depolarizing channel is well known in quantum information theory. A qubit in a pure state described by the density matrix $\rho$ has a probability $\lambda$ to depolarize (become a mixed state) and $(1-\lambda)$ to remain the same. So, the map of this evolution is    
%
\begin{equation}
  \Ecal(\rho)=\lambda \frac{\mathbb{I}}{2}+(1-\lambda)\rho.  
  \label{map}
\end{equation}

We can write the previous equation by using Pauli operators to write the map of Eq.\ref{map} in the sum-operator formalism  \cite{nielsen2000quantum}


%
\begin{equation}
\Ecal(\rho)=\left(1-\frac{3\lambda}{4}\right)\rho+\frac{\lambda}{4}\left(\sigma_{x}\rho \sigma_{x}+\sigma_{y}\rho \sigma_{y}+\sigma_{z}\rho \sigma_{z}\right).   
\end{equation}
We can find the most common Kraus operators for this channel if we consider a $\lambda'=\frac{3\lambda}{4}$, giving the map
\begin{equation}
\Ecal(\rho)=\left(1-\lambda'\right)\rho+\frac{\lambda'}{3}\left(\sigma_{x}\rho \sigma_{x}+\sigma_{y}\rho \sigma_{y}+\sigma_{z}\rho \sigma_{z}\right),   
\label{eq:map2}
\end{equation}
where we see that the Kraus operators are
\begin{align}
K_0 &= \sqrt{1-\lambda}\, I = \sqrt{1-\lambda} \begin{pmatrix} 1 & 0 \\ 0 & 1 \end{pmatrix} & K_1 &= \sqrt{\frac{\lambda}{3}}\, \sigma_x = \sqrt{\frac{\lambda}{3}} \begin{pmatrix} 0 & 1 \\ 1 & 0 \end{pmatrix} \\
K_2 &= \sqrt{\frac{\lambda}{3}}\, \sigma_y = \sqrt{\frac{\lambda}{3}} \begin{pmatrix} 0 & -i \\ i & 0 \end{pmatrix} & K_3 &= \sqrt{\frac{\lambda}{3}}\, \sigma_z = \sqrt{\frac{\lambda}{3}} \begin{pmatrix} 1 & 0 \\ 0 & -1 \end{pmatrix}.
\end{align} 
With the Kraus operators and the fact that we can write $\rho$ as 
\begin{equation}
\rho=\frac{1}{2}\left(\mathbb{I}+\Vec{r}.\Vec{\sigma}\right),    
\end{equation}
where $\Vec{r}=(r_{x},r_{y},r_{z})$ is the Bloch vector and $\Vec{\sigma}=(\sigma_{x},\sigma_{y},\sigma_{z})$ is a vector with the Pauli's matrices, we can study how the effect of the channel on the Bloch's sphere
, and we can calculate the outcome
\begin{equation}
 \sum_{a=0}^{3}K_{a}\rho K_{a}^{\dagger}=  \frac{1}{6}\begin{pmatrix} [3+(3-4\lambda)r_{z}] & (3-4\lambda)(r_{x}-ir_{y}) \\ (3-4\lambda)(r_{x}+ir_{y}) & [3-(3-4\lambda)r_{z}] .\end{pmatrix}
\label{eq:map3}
\end{equation}

\begin{figure}[h!]
\centering 
\includegraphics[scale=0.35]{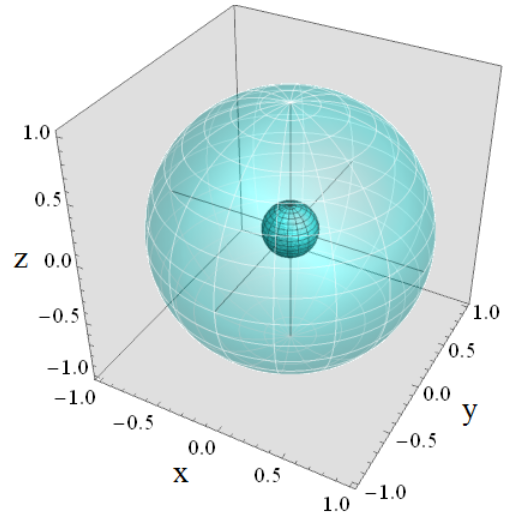}
\caption{Contraction of Bloch sphere under depolarizing channel action.}
\label{fig:Bloch-Esfera}
\end{figure}

To end, if we compare Equations (\ref{eq:map3}) and (\ref{eq:map2}), we can see that the Bloch sphere contracts uniformly as a function of $\lambda$, as shown in Eq.(\ref{eq:blochvector}) and in the Fig \ref{fig:Bloch-Esfera} as an example.

\begin{equation}
(r'_{x},r'_{y},r'_{z})\rightarrow \left(\left(1-\frac{4\lambda}{3}\right)r_{x},\left(1-\frac{4\lambda}{3}\right)r_{y},\left(1-\frac{4\lambda}{3}\right)r_{z}\right)
\label{eq:blochvector}
\end{equation}
\subsection{Spin-Orbit modes}

    In order to emulate the interaction between a system and the environment experimentally, we will consider the degrees of freedom of light, the polarization, and the Hermite-Gaussian transverse modes. It is known that polarization are related to the vector nature of the electromagnetic field, where we can write any polarization state considering the horizontal ($\hat{e}_{H}$) and vertical ($\hat{e}_{V}$) basis $\{\hat{e}_{H},\hat{e}_{V}\}\equiv \{H, V\}$, and for the HG modes, we can have a basis with the first order modes considering the spatial function of the HG oriented along the horizontal  ($\psi_{h}$) and vertical ($\psi_{h}$) directions $\{\psi_{h},\psi_{v}\} \equiv \{h, v\}$. Therefore, these two degrees of freedom can be addressed independently, and a combined mode basis can be built from a tensor product of these two bases, and a beam with these two degrees are what we call Spin-Orbit (SO) modes. As discussed in Ref.\cite{Pereira}, the quantization of the electromagnetic field in this basis leads to a bipartite system of internal degrees of freedom of a single photon. Considering an intense laser beam can be described as a coherent state with a macroscopic number of photons, we use Dirac notation to describe the DoF of the most general SO modes as  \cite{Pereira}
%
%
\begin{equation}
\ket{\Psi_{SO}}=A_{Hh}\ket{Hh}+A_{Hv}\ket{Hv}+A_{Vh}\ket{Vh}+A_{Vv}\ket{Vv}.
\label{eq:sogeneral}
\end{equation}
where $A_{ij}$ is the amplitude and $\sum_{i,j}|A_{ij}|^{2}=1$ , with $i=H,V$ and $i=h,v$ being the polarization and transverse mode indexes. We can notice from Equation \ref{eq:sogeneral} that these modes can be separable and non-separable, i.e, we can factorized in product of transverse mode and polarization
\begin{equation}
\ket{\Psi_{SO}}=(A_{h}\ket{h}+A_{v}\ket{v})\otimes(A_{H}\ket{H}+A_{V}\ket{V}),
\end{equation}
where we used $A_{ij}=A_iA_j$. In this case, we say the SO is separable. For any other situation, the SO mode is called non-separable mode. The quantization of the electromagnetic field in such a basis lead to entanglement between the internal degree of freedom of light \cite{Pereira}. Note that Eq.\ref{eq:sogeneral} presents the same structure of a general bipartite system. Then, to 
characterize the spin-orbit separability with a definition similar to concurrence
\begin{equation}
 C= 2 | A_{Hh} A_{Vv} - A_{Hv}A_{Vh}| ,  
\end{equation}
so, if $C=0$, we have a separable mode, but if $0<C\leq 1$ is non-separable. The case $C=1$ indicates a maximally non-separable mode and can be written as
\begin{equation}
\begin{aligned}
\ket{\Phi_{\pm}}=\frac{\ket{Hh}\pm \ket{Vv}}{\sqrt{2}},  \\
\ket{\Psi_{\pm}} = \frac{\ket{Hv}\pm \ket{Vh}}{\sqrt{2}},
\end{aligned}
\label{eq:mnsmode}
\end{equation}
that share the same mathematical structure as the maximally entangled Bell states. These maximally non-separable modes allow us to study entanglement and quantum information with an intense laser beam and even with photons \cite{topo, bell1, bell3}. A complete study of the partial separability of these modes can be seen in ref. \cite{lamego2023partial}.

\subsection{Solvay-Kitaev decomposition}

\begin{figure}[h!]
\centering 
\includegraphics[scale=0.8]{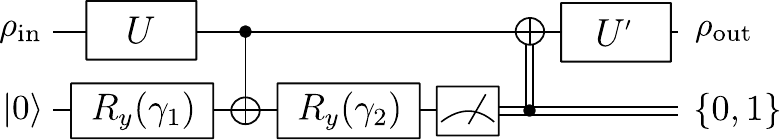}
\caption{Logical circuit for implementing an arbitrary quantum channel $\Ecal$, as defined in Eq.~(\ref{qechannel}), acting on a single-qubit system. Note that the present circuit implements the channels $\Ecal_a^e$ and $\Ecal_b^e$ individually  \cite{PhysRevLett.111.130504}.}
\label{fig:circuit}
\end{figure}

A few years ago, inspired by the Solovay-Kitaev theorem, it was demonstrated that any single-qubit CPTP channel $\Ecal$ can be decomposed into a convex combination as 
\begin{equation}
\label{qechannel}
\Ecal = p \Ecal^e_a + (1-p)\Ecal^e_b,
\end{equation}
where $0 \leq p \leq 1$ and $\Ecal^e_i,\,\, i=a,b$ are two quasi-extreme channels \cite{PhysRevLett.111.130504}. Such a decomposition can be implemented using one ancillary qubit, two CNOT gates, and four single-qubit operations, as shown in the logical circuit in Fig. \ref{fig:circuit} and described in Ref.~\cite{PhysRevLett.111.130504}. Unitary operators $U$ and $U^\prime$ belong to the SU(2) group and will assume different forms depending on the channel to be emulated with the SK-decomposition.

As detailed in Ref.\cite{PhysRevLett.111.130504}, 
$\Ecal^e_i (i=a,b)$ are represented in terms of modified Kraus operators 
\begin{equation}
M_i = U K_iU^\prime.
\end{equation}
The map $\Ecal \equiv \Ecal(\lambda, \alpha, \beta, \gamma_1, \gamma_2, p) $ where the parameters $\lambda, \alpha, \beta$ are variables of the kraus operator, being $\lambda$ associated to the evolution within the channel. $\gamma_1$ and $\gamma_2$ are angles related to rotation operations applied to the ancilla, while $p$ is the weight associated with the convex combination of the quasi-extreme channels $\Ecal^e_{a,b}$.

Our group implemented the SK-decomposition experimentally by using spin-orbit modes \cite{passos2020spin}. The system was codified in polarization DoF, while the ancilla was codified in the first-order Hermite-Gauss (HG) modes. For further details on the experimental implementation, see Ref. \cite{passos2020spin}, where the optical circuit is presented. In the experimental study referred to, the operators $R_y(\gamma)$ that acted in the HG mode are performed by a sequence of two Dove Prisms (DP), one horizontally oriented and the other rotated by $\gamma/2$. On the other hand, $U$ and $U^\prime$ act on the polarization DoF (system) and are performed by a sequence of a Quarter wave plate, a Half wave-plate, and another Quarter wave-plate. Using such a combination of wave plates enables us to build a general unitary transformation on the SU(2) group. The $U_{DP}$ operator is giving by
\begin{equation}
    U_{DP}= QWP(0)HWP(\pi/2)QWP(-\pi/2).
\end{equation}
All relations among the parameters of the SK-decomposition and the optical elements are established in detail in Ref.\cite{passos2020spin}. To conclude, the parameters required for the experimental realization of the depolarizing channel are listed in Table \ref{Tab:DPchannel2}.


 %
\begin{table}[h!]
\centering

\begin{tabular}{|c|c|c|c|c|c|c|c|c|c|c|c|c|c|}
\hline
\multirow{2}{*}{$\lambda$} & \multicolumn{6}{c|}{$\Ecal^e_1$}                                                                          & \multicolumn{7}{c|}{$\Ecal^e_2$}                                                                             \\ \cline{2-14} 
                           & $\alpha$        & $\beta$         & $2\gamma_1$     & $2\gamma_2$      & $U$ & $U^\prime$ & $\alpha$         & $\beta$         & $2\gamma_1$     & $2\gamma_2$ & $U$ & $U^\prime$ & p    \\ \hline
0                          & 0               & 0               & $\frac{\pi}{2}$ & -$\frac{\pi}{2}$ & none           & none            & $\frac{\pi}{4}$ & $\frac{\pi}{4}$ & $\frac{\pi}{2}$ & 0           & $U_{\text{DP}}$       & none            & 1    \\ \hline
0.31                       & $\frac{\pi}{3}$ & $\frac{\pi}{3}$ & $\frac{\pi}{2}$ & $\frac{\pi}{6}$  & none           & none            & $\frac{\pi}{4}$ & $\frac{\pi}{4}$ & $\frac{\pi}{2}$ & 0           & $U_{\text{DP}}$       & none            & 0.76 \\ \hline
0.5                        & $\frac{\pi}{6}$ & $\frac{\pi}{6}$ & $\frac{\pi}{2}$ & -$\frac{\pi}{6}$ & none           & none            & $\frac{\pi}{4}$ & $\frac{\pi}{4}$ & $\frac{\pi}{2}$ & 0           & $U_{\text{DP}}$       & none            & 0.66 \\ \hline
0.75                       & $\frac{\pi}{4}$ & $\frac{\pi}{4}$ & $\frac{\pi}{2}$ & 0                & none           & none            & $\frac{\pi}{4}$ & $\frac{\pi}{4}$ & $\frac{\pi}{2}$ & 0           & $U_{\text{DP}}$       & none            & 0.5  \\ \hline
1                          & $\frac{\pi}{2}$ & $\frac{\pi}{2}$ & $\frac{\pi}{2}$ & $\frac{\pi}{2}$  & none           & none            & $\frac{\pi}{4}$ & $\frac{\pi}{4}$ & $\frac{\pi}{2}$ & 0           & $U_{\text{DP}}$       & none            & 0.33 \\ \hline

\end{tabular}
\caption{Parameters to implement the Depolarizing channel in spin-orbit SK-decomposition. $U_{\text{DP}}$ is the operator built to Depolarizing channel implementation. } 
\label{Tab:DPchannel2}
\end{table}
%



\subsection{ Quantum Coherence}\label{sec:QuantumCoh}

Let us assume an arbitrary d-dimensional pure quantum state $\rho$ described in a Hilbert space $\Hcal$. The general form of an incoherent quantum state is written as 
\begin{equation}
 \delta =\sum_{k=0}^{d-1} p_k \ket{k}\bra{k} ,
\end{equation}
where 0 $\leq p_k \leq 1$, with $\sum_k p_k = 1$, and $\ket{k}$ being an orthonormal reference basis. A proper measure of quantum coherence $C(\rho)$ is defined as a nonnegative functional that vanishes on the set of incoherent states and does not increase under incoherent completely positive and trace-preserving (ICPTP) operations \cite{Baumgratz2014}. One of the classes of coherence measurement is described  in terms of the distance (or pseudo-distance) $\Dcal$ between the quantum state $\rho$ and the closest incoherent state $\delta_{min}$, in such a way that
\begin{equation}
 C(\rho) = \Dcal(\rho,\delta_{min}).
\end{equation}
Within the distance based quantum coherence measures, one of the most well-known is the coherence $l_1$-norm, which is defined as $C(\rho) = || \rho -\delta_{min} ||_{l_1}$, where $||O||_{l_1} = \sum_{k \neq k^{\prime}} |\bra{k}O\ket{k^{\prime}}|$ denotes the $l_1$-norm of the operator $O$ \cite{Baumgratz2014}. In this regard, the $C_{l_1}$ takes the follwoing form
\begin{equation}
     C_{l_1}(\rho)=\sum_{k,k^{'},k\neq k^{'}} \left|\bra{k}\rho \ket{k'} \right|,
\end{equation}
%
In another way, if one assumes the density operator formalism for the case of one-qubit ($d=2$) state, where $\rho=\left(I+\vec{r}\dot{.}\vec{\sigma}\right)/2$, with $\vec{r}=\left(r_x, r_y, r_z\right)$ being the Bloch vector and $\vec{\sigma}=\left(\sigma_x, \sigma_y, \sigma_z\right)$ representing the vector composed by Pauli matrices and $I$ denoting the identity operator, the $l_1$-norm quantum coherence becomes
\begin{equation}
     C_{l_1}(\rho) \equiv C_{l_1}(r_x, r_y) = \sqrt{r^2_x + r^2_y}.
\end{equation}

However, the $l_1$-norm quantum coherence is a basis-dependent concept, where even local unitary operations could change the quantum coherence of a given state. In this regard, a basis-free measure of quantum coherence can be defined if one performs a maximization over all \textit{local} unitary transformations \cite{Yu2016Sep,Streltsov2018May} as follows
\begin{equation}\label{C: MAX}
C_{\mathrm{\text{max}}}(\rho) = C\left(U_{\text{max}}\; \rho\; U^\dagger_{\text{max}}\right),
\end{equation}
where $U_{\text{max}}$ describes a \textit{local} unitary operation that maximizes $C(\rho)$. By assuming the operator $U_{max} = \sum_n \ket{n_+}\bra{\psi_n}$, where $\{\ket{\psi_n}\}$ is the eigenstate of the quantum state $\rho$ and $\{\ket{n_+}\}$ is a mutually unbiased basis with respected to the incoherent basis $\{\ket{k}\}$ \cite{Yu2016Sep}, the maximal quantum coherence takes the form 
\begin{equation}
C_{max}(\rho) = \sum_{k,k^{\prime}}|\langle k\vert \rho -I/d \vert k^{\prime}\rangle|,
\end{equation}
or, in terms of the Bloch sphere formalism,
\begin{equation}
 C_{max}(\rho) \equiv C_{max}(r) = r~.
\end{equation}

In this work, beyond performing polarization state reconstruction via quantum tomography, we employ the $l_1$-norm ($C_{l_1}$) and the maximal quantum coherence ($C_{max}$) as figures of merit to quantify the effects of decoherence on a quantum state subjected to a depolarizing channel.



%

%


\section{Proposal of compact circuit for Depolarizing Channel simulation} \label{sec:CompCircuit}

\begin{figure}[H]
\centering 
\includegraphics[scale=0.45]{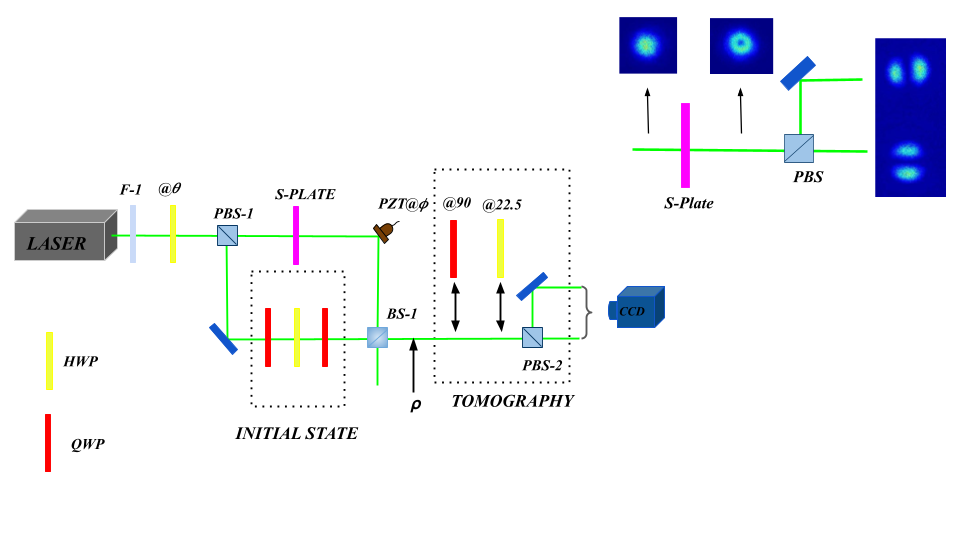}
\caption{Experimental setup. S-plate prepares a maximally non-separable state, that presents maximally mixed polarization state by tracing transverse modes. The action of the S-plate in a Gaussian beam horizontally polarized is showed in the image at superior right part of the figure.}  
\label{fig:compact-depolarizing-channel}
\end{figure}

The compact circuit that simulates the evolution of a quantum state under effect of a Depolarizing channel is presented in the Fig.\ref{fig:compact-depolarizing-channel}. In the same way as the SK-decomposition, our quantum state is described by a single qubit codified in the polarization DoF ($\ket{0}\equiv\ket{H}$ and $\ket{1}\equiv\ket{V}$) and the ancillary is codified in the transversal Gaussian ($HG_{00} \equiv \ket{G}$) and Hermite-Gauss ($HG_{10} \equiv \ket{h}$ and $HG_{01} \equiv \ket{v}$) modes.

We begin with a vertically polarized laser beam (532 nm) passing through a filter (F-1) to attenuate its intensity. Therefore, we can write the initial state of the laser as   
%
\begin{equation}
 \ket{\psi^{0}}=\ket{V}\otimes\ket{G}=\ket{VG}, 
\end{equation}
afterwards, a Half Wave Plate (HWP) with an angle $\textbf{@}\theta$ with respect to the horizontal acts on the state $\ket{\psi^{0}}$, generating the state
%
\begin{equation}
\ket{\psi^{1}}=\sin(\theta)\ket{H}\otimes\ket{G}-\cos{(\theta)}\ket{V}\otimes\ket{G}.  
\end{equation}

The depolarizing parameter $\lambda$, defined in Eq.~\ref{map}, is experimentally controlled by the HWP$_1$ angle $\theta$.


The next step is to position a polarized beam splitter (PBS-1), where the horizontal component is going to be transmitted, leaving the state to be 
\begin{equation}
 \ket{\psi^{2}}=\sin(\theta)\ket{H}\otimes\ket{G}   
\end{equation}
in the transmitted arm and the vertical one is reflected to the order arm with the state
\begin{equation}
 \ket{\psi^{3}}=-\cos{(\theta)}\ket{V}\otimes\ket{G}.  
\end{equation}



To prepare the state whose evolution in the Depolarizing channel we wish to simulate, that we call initial state, as is already well known, we can use a sequence of three wave plates in the arm of the reflected beam: a Quarter Wave Plate, a Half Wave Plate, and a another Quarter Wave Plate. This set can perform arbitrary transformation in polarization degree of freedom, that corresponds to the preparation an arbitrary initial qubit. In this way, the we have the transformation $\ket{V} \rightarrow \alpha \ket{H} + \beta \ket{V}$, where $\alpha$ and $\beta$ are complex amplitudes and $|\alpha|^{2}+|\beta|^{2}=1$. The state in this arm becomes  
\begin{equation}\label{Eq.InPolState}
 \ket{\psi^{4}}=-\cos{2\theta}(\alpha \ket{H}+\beta \ket{V})\otimes \ket{G}.   
\end{equation}
The states that we are going to study are the vertical, $\ket{V}, \text{which implies}$ $\alpha=0~ \text{and}~ \beta=1$, and diagonal, $\ket{+}=\frac{\ket{H}+\ket{V}}{\sqrt{2}}, \text{which implies}$ $\alpha=\beta=\frac{1}{\sqrt{2}}$. 

Now, we will focus on the transmitted beam.  After the PBS-1, it passes through an S-wave plate, an optical element that can change the horizontal Gaussian beam into the maximal non-separable state $\frac{1}{\sqrt{2}}(\ket{Hh}-\ket{Vv})$, leading to 
\begin{equation}
\ket{\psi^{5}}=\sin{(2\theta)}\frac{1}{\sqrt{2}}(\ket{Hh}-\ket{Vv}).    
\end{equation}
The action of the S-plate in a
Gaussian beam horizontally polarized is showed in the image at superior right part of the Figure \ref{fig:circuit}. A donuts shape is observed after S-Plate. A projective measurement in the polarization degree fo freedom with a PBS separates $h$ and $v$ transverse modes.    

After the S-plate, a mirror with Piezoeletric ceramic (PZT) gives a phase $\phi$ between the paths. Moreover, to end the evolved state in Depolarizing Channel, the beams are put together in a beam splitter (BS-1), generating the state
\begin{equation}
\ket{\psi^{6}}=-cos{(2\theta)}(\alpha \ket{H}+\beta \ket{V})\otimes \ket{G}+e^{i\phi}\sin{(2\theta)}\frac{1}{\sqrt{2}}(\ket{Hh}-\ket{Vv}).
\label{eq:final_experimental-state}
\end{equation}




With our final state, we need to characterize it, this process is done by polarization state  tomography by defining 
the computational basis $(\{\ket{H},\ket{V}\})$, the diagonal/anti diagonal $(\{\ket{+},\ket{-}\})$, and the left/right circular $(\ket{L},\ket{R})$ polarization basis through a PBS, a Quarter-Wave Plate ($@90^\circ$), and a Half-Wave Plate ($@22.5^\circ$), alternately. The PBS outputs are projected onto a screen and the images are recorded by a CCD camera in a unique frame. The normalized intensities $\mathcal{I_j} = I_{j}/I_T$, with $j= H,V,+,-,L,R$, being $I_T$ the total intensity, play the role of probabilities to calculate the Stokes parameters and reconstruct the density matrices of the evolved states. This process alow us to only have the information of the polarization qubit, similar to a partial trace.  Hence, if we take the density operator of Eq.(\ref{eq:final_experimental-state}) $\rho = \ket{\psi^6}\bra{\psi^6}$, we can obtain the reduced density matrix of the polarization degree of freedom
\begin{equation}
\begin{split}
&\rho^{Pol}=\cos^{2}{(2\theta)}\left(|\alpha|^{2}\ket{H}\bra{H}+\alpha \beta^{*}\ket{H}\bra{V}+\beta \alpha^{*}\ket{V}\bra{H}+|\beta|^{2}\ket{V}\bra{V}\right)+\\
&+\frac{\sin^{2}{(2\theta)}}{2}\left(\ket{H}\bra{H}+\ket{V}\bra{V}\right),\\  
\end{split}
\label{polfinal}
\end{equation}
which is analogous to the expected final state after evolution in the Depolarizing channel. Note that for $\theta = 0^\circ$, which corresponds to $\lambda=0$, we have the initial state for polarization given by Eq.\ref{Eq.InPolState}. For $\theta = 45^\circ$ ($\lambda=1$), we have the maximally mixed state for polarization as expected for $t\rightarrow \infty$. So if we consider the polarization space, we can simulate the depolarization of one qubit.

\section{Experimental results}

In this section, we present the experimental results for the depolarizing channel based on the SK decomposition and its simulation implemented using the compact optical circuit described in Sec.~\ref{sec:CompCircuit}. To illustrate the effect of Depolarization on a qubit evolving under the depolarizing channel, we consider two input states: (i) a vertically polarized state, $\rho=\ket{V}\bra{V}$, and ii) a coherent superposition state, $\rho=\ket{+}\bra{+}$.


\subsection{Results for the Solovay-Kitaev decomposition}

\begin{figure}[H]
\centering 
\includegraphics[scale=0.32]{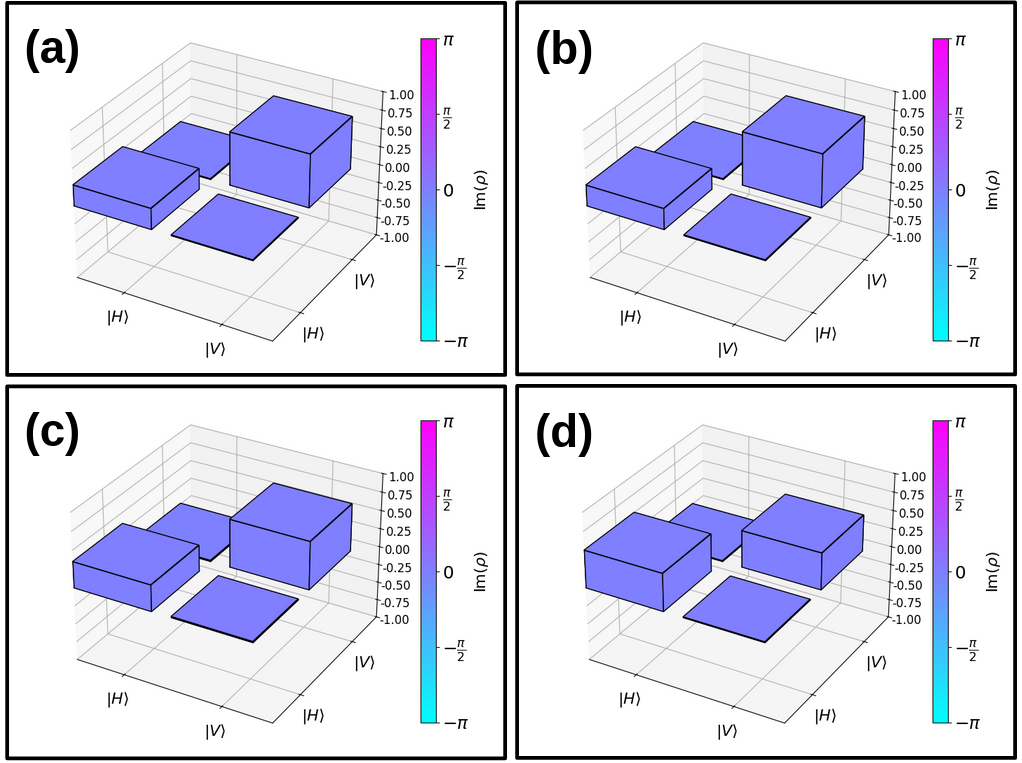}
\caption{Density Matrices of the state $\rho=\ket{V}\bra{V}$ under the evolution in the depolarizing channel implemented by the Solovay-Kitaev decomposition. $\lambda'$ is the depolarization parameter. The panels correspond to (a) $\lambda'=0$, (b) $\lambda'=0.25
$, (c) $\lambda'=0.5$, and (d) $\lambda=0.75$.}
\label{fig:Vertical_Density_Marcello}
\end{figure}

Figure \ref{fig:Vertical_Density_Marcello} presents the experimental reconstructed density matrices, via polarization quantum tomography, of the initial quantum state $\ket{V}\bra{V}$ under the effect of a depolarizing quantum channel. In panel (a), we present the result for $\lambda^\prime=0$, i.e, the initial state unaffected by the channel. In this case, it is expected that the matrix element $\ket{H}\bra{H}$ should be null, however due to experimental error occasioned by the optics devices, it is straightforward to notice that we have an small contribution of such polarization component after the initial state undergoes through the quantum channel. Such an effect is testified by a lower value of the fidelity, $F=71.86\%$. We strongly believe that this effect can be explained by the fact that the Dove prism slightly disturbs the state, leading to a small $H-$polarization component \cite{SolovayKitaevPassos}. Such an effect is also expected to happen in the case of a diagonal polarized quantum state. However, as the $H-$polarization component is already present with a significant contribution to the whole state, the relative deviation is less pronounced. It is important to notice that this effect is scaled up by the presence of four Dove prisms in the SK-decomposition optical setup. As expected, the state evolution under the channel leads to partially mixed states (b) and (c). The final state becomes totally mixed for $\lambda^\prime=0.75$, which corresponds to a sufficiently long time ($\lambda=1$), as illustrated in Fig. (d). For this state, the Fidelity is $F=99.94\%$. It is interesting to highlight that the action of the depolarizing channel in our quantum state $\ket{V}\bra{V}$ does not create any coherence elements in the density matrix during its evolution, as expected.

\begin{figure}[H]
\centering 
\includegraphics[scale=0.32]{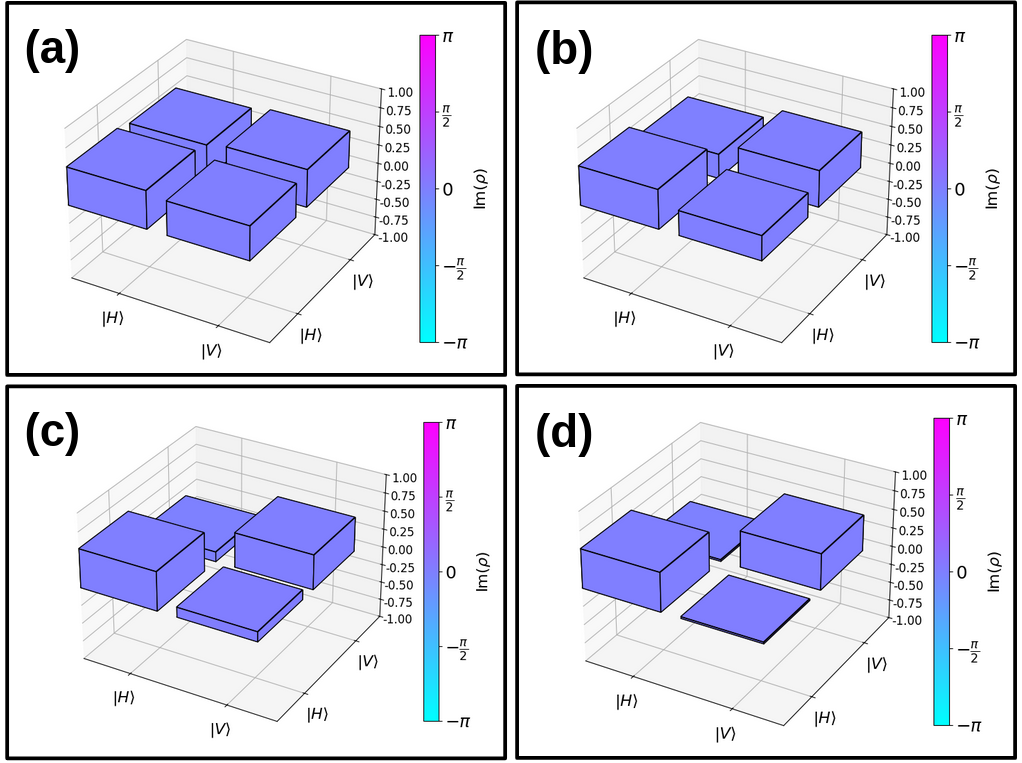}
\caption{Density Matrices of the state $\rho=\ket{+}\bra{+}$ under the evolution in the depolarizing channel implemented by the Solovay-Kitaev decomposition. $\lambda'$ is the depolarization parameter. The panels correspond to (a) $\lambda'=0$, (b) $\lambda'=0.25
$, (c) $\lambda'=0.5$, and (d) $\lambda'=0.75$.}
\label{fig:Diagonal_Density_Marcello}
\end{figure}

Figure \ref{fig:Diagonal_Density_Marcello} presents the density matrices of the second quantum state under analysis, $\ket{+}\bra{+}$.
As before, panel (a) is associated with $\lambda^\prime=0$ and hence no evolution through the quantum channel. In this case, the Fidelity of the experimentally reconstructed quantum state is equal to 
$F=94.78\%$. It is worth noticing that the maximum coherent quantum state $\ket{+}\bra{+}$ starts to lose its coherence terms (non-diagonal elements in the respective density matrix) as long as the interaction with the quantum channel increases. In the asymptotical time evolution (long time interaction), the maximally mixed state is achieved with Fidelity $F=99.82\%$ (d).

Even though the reconstructed density matrices already let us observe the effect of the depolarization of the initial quantum state, a stronger figure of merit is needed. To this end, we analyze both quantum coherence described in Sec \ref{sec:QuantumCoh}.

\begin{figure}[H]
\centering 
\includegraphics[scale=0.32]{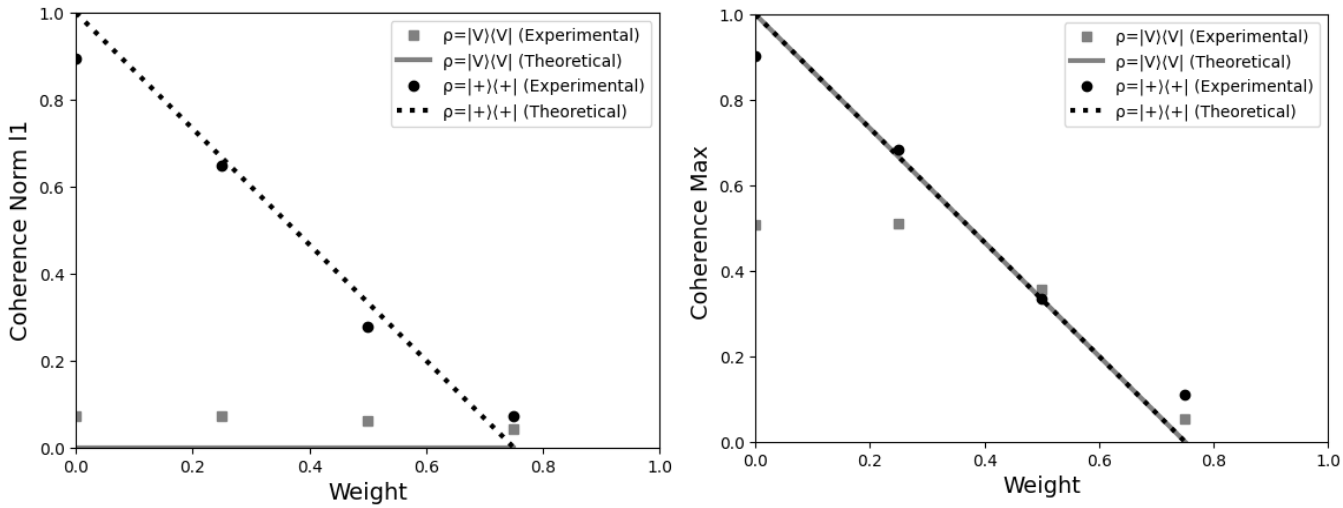}
\caption{$l_1$-norm Coherence (left) and Maximal Coherence (right) behavior for the Depolarizing Channel with Solovay-kitaev decomposition.}
\label{fig:Coherence_Marcello}
\end{figure}

Figure \ref{fig:Coherence_Marcello} presents the experimental results as well as the theoretical predictions for $l_1-$norm Coherence (left) and Maximal Coherence (right) for both initial quantum states under study. Since the vertical polarization state does not have any
coherence elements, the $l_1-$norm coherence is expected to be null, as illustrated by the theoretical gray solid line. The squared gray experimental points clearly reproduce this behavior, even though a slight deviation from the expected value (equal to zero) is observed due to the experimental limitations.
On the other hand, for the Diagonal state, the situation changes. Since this quantum state has coherence elements which degrade as soon as the interaction with the quantum channel increases, the $l_1-$norm coherence has its values monotonically decreasing from 1 (maximum) to zero (minimum - completely incoherent). The theoretical expectation of this behavior is illustrated by the black dashed line. It is straightforward to see that our experimental results (black points) are in very good agreement with the theoretical predictions.

In the case of the maximal Coherence ($C_{max}$), as this coherence can also be seen as a measure of purity, it is expected that the $C_{max}$ starts with the value equal to 1 and monotonically decrease to zero as soon as both initial pure quantum states become mixed. This dynamic is described by the theoretical predictions for the state $\ket{V}\bra{V}$ (gray solid line) and $\ket{+}\bra{+}$ (black dashed line). It is easy to see that the experimental results for both states are in good agreement with the theoretical predictions, except for the case $\lambda^\prime=0$, which presents the worst fidelity in the density matrix reconstruction as previously discussed.

To conclude, it is worth summarizing the difference in the results presented here for both quantum coherence. For instance, for the initial state $\ket{V}\bra{V}$, the $l_1$-norm coherence is equal to zero because the state is diagonal in the reference basis and thus contains no basis-dependent superposition, whereas the maximal coherence decreases monotonically under depolarization as it quantifies the state’s purity, which is progressively reduced by mixing with the maximally mixed state. The comparison of both coherence for the diagonal polarization state can be explained in the same way.

\subsection{Results for the proposed compact circuit}
In this subsection, we will present our experimental results from the proposed compact circuit, which simulates the effect of a depolarizing channel. For the purpose of comparing our circuit with the Solovay-Kitaev experimental proposal, we considered the same states, i.e, $\rho=\ket{V}\bra{V}$ and  $\rho=\ket{+}\bra{+}$. The first important aspect to mention is the facility to align and perform a complete measurement. Another important point to mention is that here we directly emulate the weight $\lambda$.  

\begin{figure}[H]
\centering 
\includegraphics[scale=0.4]{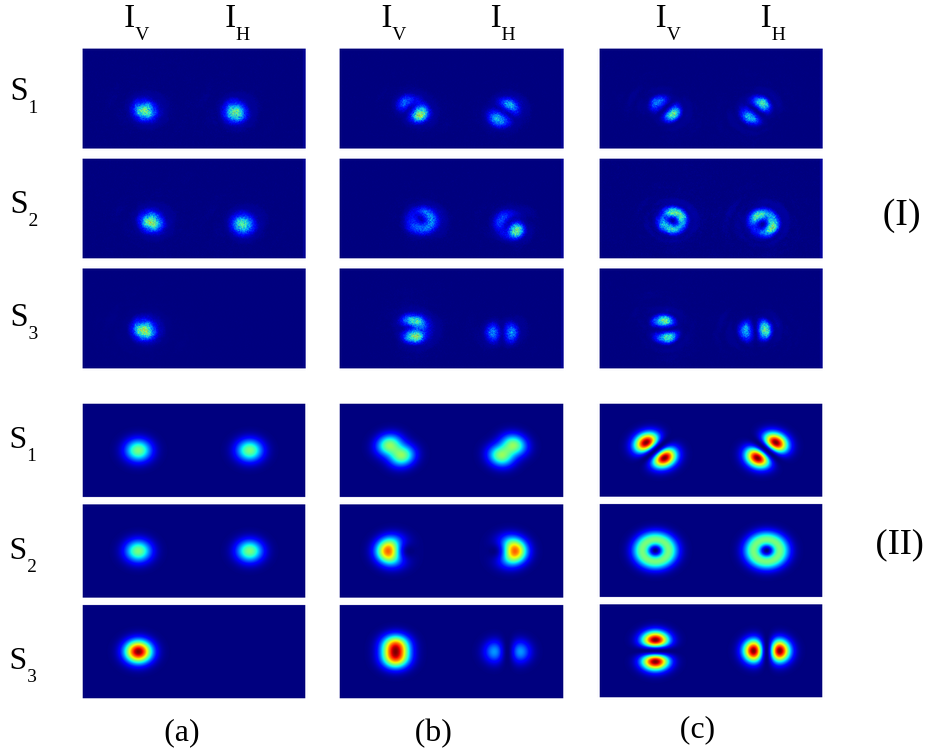}
\caption{Tomography results for the state $\rho=\ket{V}\bra{V}$ with (a) $\theta=0$, (b) $\theta=30$, (c) $\theta=45$. (I) are the experimental results, and (II) are the theoretical predictions.}
\label{fig:Vertical_Tomography}
\end{figure}

\begin{figure}[H]
\centering 
\includegraphics[scale=0.4]{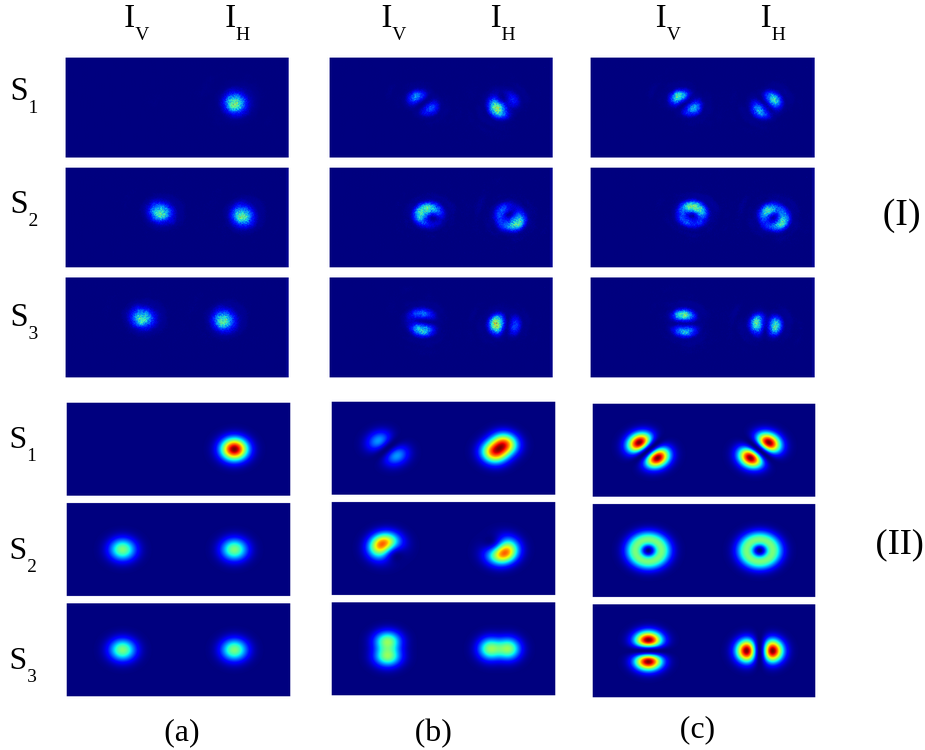}
\caption{Tomography results for the state $\rho=\ket{+}\bra{+}$ with (a) $\theta=0$, (b) $\theta=30$, (c) $\theta=45$. (I) are the experimental results, and (II) are the theoretical predictions.}
\label{fig:Diagonal_Tomography}
\end{figure}

Figures \ref{fig:Vertical_Tomography} and \ref{fig:Diagonal_Tomography} present the false color output images of the polarization tomography for the states $\rho=\ket{V}\bra{V}$ and $\rho=\ket{+}\bra{+}$, respectively. The group of images (I) corresponds to the experimental results, and the group (II) to the numerical simulation. Furthermore, $I_{i} (i=H,V)$ indicates the intensity output that is transmitted ($I_{H}$) and reflected ($I_{V}$) by the PBS$_2$ in the tomography process, and $S_{i}$ (i=1,2,3) indicates the calculated Stokes parameters.

In both cases, we have a very good agreement between theory and experiment. A small difference in the mode shape during the intermediate evolution—Figs. \ref{fig:Vertical_Tomography}(b) and \ref{fig:Diagonal_Tomography}(b)—can be observed. This discrepancy is mainly due to slight misalignment during the interference process. However, it does not affect the results, since determining the polarization state density matrix only requires the normalized intensity of the respective state projected by PBS$_2$ to calculate the Stokes parameters and reconstruct the state.



\begin{figure}[H]
\centering 
\includegraphics[scale=0.32]{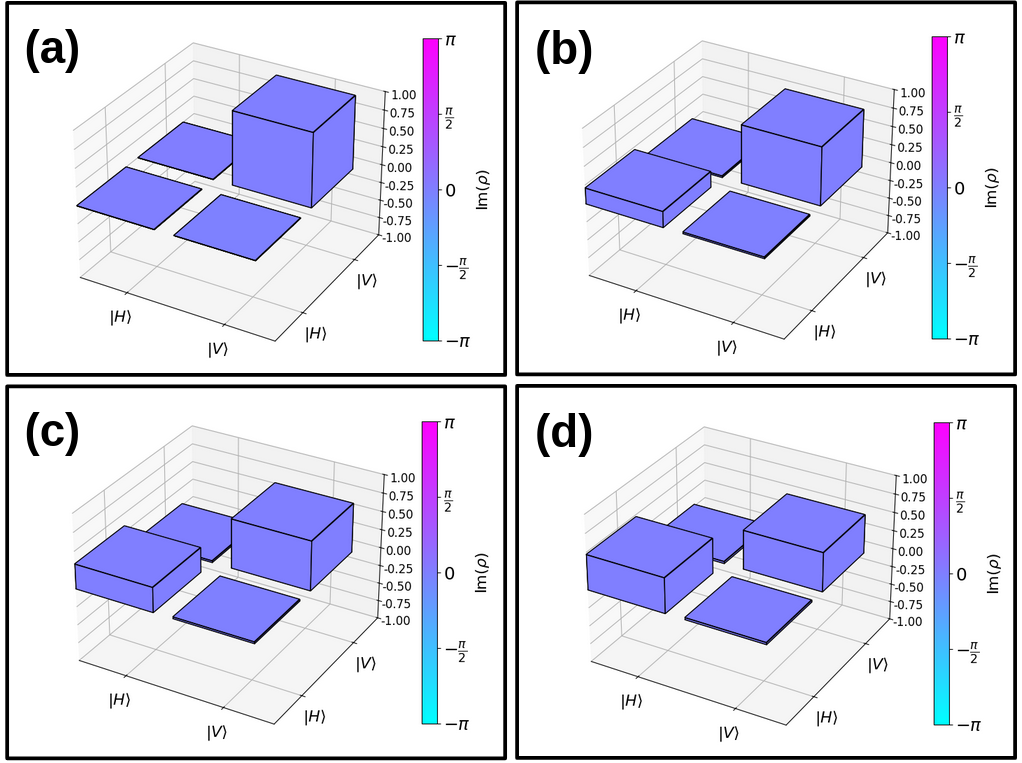}
\caption{Density Matrices of the state $\rho=\ket{V}\bra{V}$ obtained from the compact circuit. $\lambda$ is the depolarization parameter. The panels correspond to (a) $\lambda=0$, (b) $\lambda=0.40$, (c) $\lambda=0.75$, and (d) $\lambda=1$.}
\label{fig:Vertical_Density_Gabe}
\end{figure}

\begin{figure}[H]
\centering 
\includegraphics[scale=0.32]{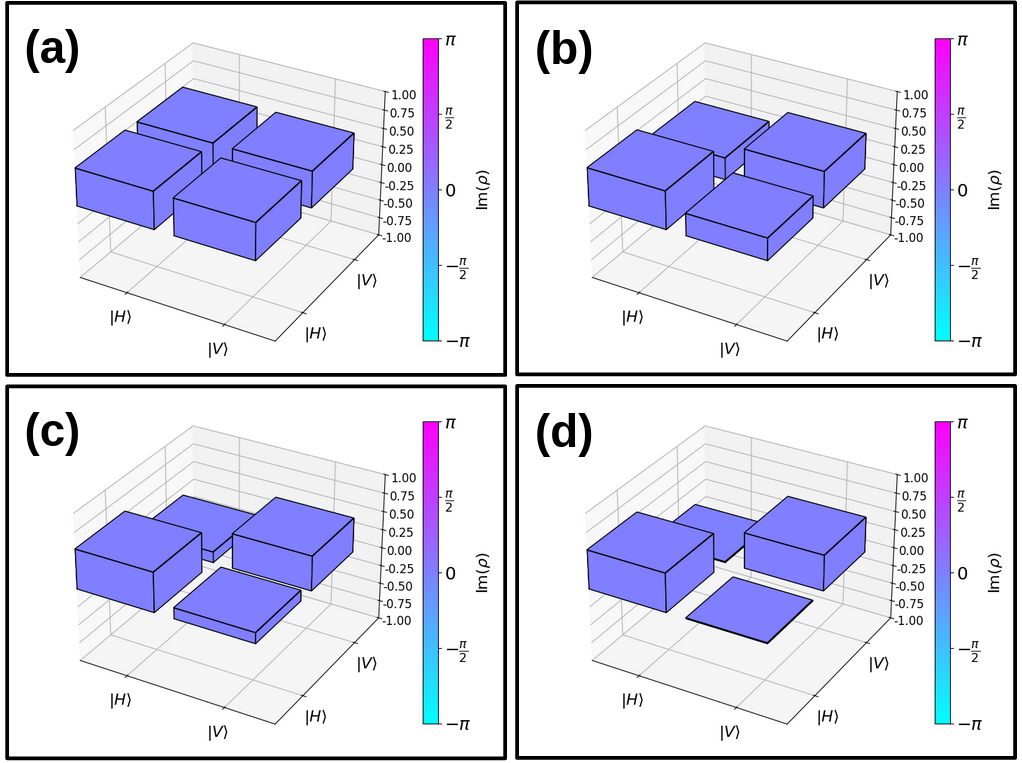}
\caption{Density Matrices of the state $\rho=\ket{+}\bra{+}$ obtained from the compact circuit.$\lambda$ is the depolarization parameter. The panels correspond to (a) $\lambda=0$, (b) $\lambda=0.40$, (c) $\lambda=0.75$, and (d) $\lambda=1$.}
\label{fig:Diagonal_Density_Gabe}
\end{figure}

Figure \ref{fig:Vertical_Density_Gabe} shows the density matrix for the vertical state. A clear improvement of the results is observed. For $\lambda=0$, the Fidelity was $F=100\%$ (a). The evolution arrived at the maximally mixed state (d) with $F=99.83\%$. For the diagonal state $\rho=\ket{+}\bra{+}$, Figure \ref{fig:Diagonal_Density_Gabe} shows the results also with excellent fidelity. For the final state, maximally mixed, we have the Fidelity equal to $99.91\%$.


\begin{figure}[H]
\centering 
\includegraphics[scale=0.32]{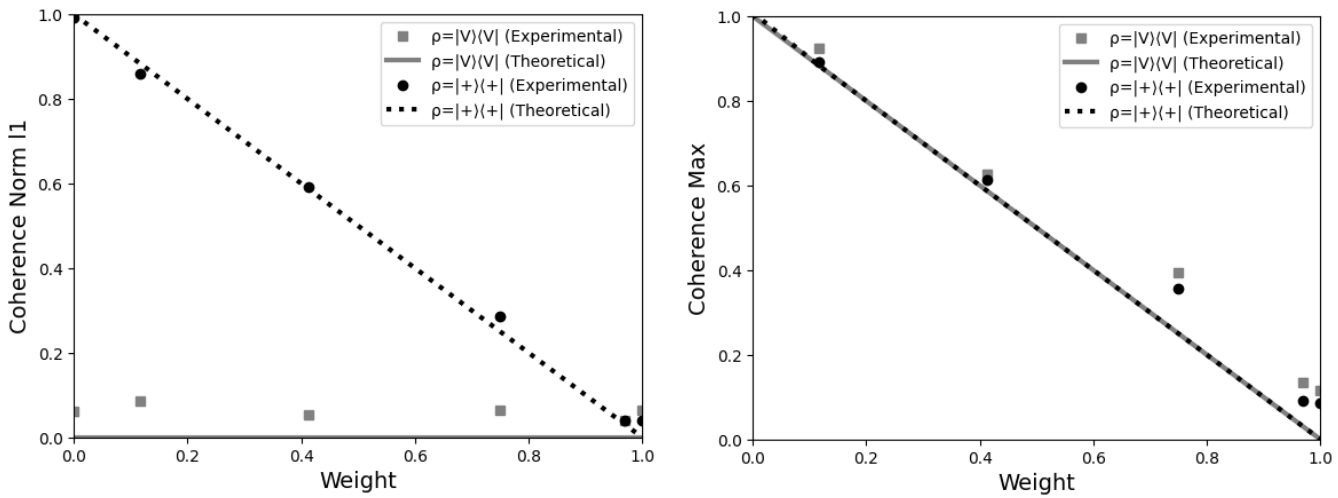}
\caption{$l_1$-norm Coherence (left) and Maximal Coherence (right) behavior for the simulation of Depolarizing channel with the proposed compact circuit.}
\label{fig:Coherence_Gabe}
\end{figure}

The results for both coherence measures are presented in Fig. \ref{fig:Coherence_Gabe} and follow the same structure than the one previously discussed. Firstly, the quantum coherence for both initial states under the compact circuit evolution reproduces the same behavior obtained with the SK-decomposition experimental proposal. This behavior is consistent with the reconstructed density matrices presented previously, allowing us to confirm that our compact circuit indeed reproduces the dynamics of these states under depolarization. Moreover, the experimental results are in excellent agreement with the corresponding theoretical predictions. Finally, it is worth highlighting the improvement in the experimental results for $C_{\max}$ of the state $\ket{V}\bra{V}$ compared to the SK-decomposition scheme. This result illustrates the potential of our compact optical circuit for applications in quantum information, particularly for the generation of different mixed polarization states.

\section{Conclusions}

In conclusion, we present a robust simulation of the Depolarizing channel by using two strategies exploring spin-orbit modes. The first one is based on the Solovay-Kitaev decomposition, extending the proposal of Ref.\cite{SolovayKitaevPassos}. The results are in good agreement with the expectations of Quantum Information Theory. The second one is a proposal and experimental implementation of a compact linear optical circuit to emulate an arbitrary state under Depolarizing channel effect. The power of this second experimental implementation 
is not just to increase the dimension of the Hilbert space of the system 
by making use of an ancillary system, as done with the SK-decomposition. Here, the key procedure is to consider a maximally non-separable mode, which enables the generation of a maximally mixed state when one takes the partial trace in the ancillary system. In this way, we can perform a convex sum with the state of the system whose evolution we wish to study.

 The calculations of the unitary transformations of the circuit in the initial states for the compact circuit, showed that the partial trace in the transverse mode is in good agreement with experimental results, thus reinforcing the applicability of our method. The results of the compact circuit are noticeably more robust than those produced by the spin-orbit Solovay-Kitaev decomposition circuit. On top of it, the simplicity of the proposed optical circuit opens an avenue to investigate quantum information protocols that make use of partially and/or maximally mixed states.
 The ability to generate mixed states with tunable depolarization enables controlled studies of robustness, noise-induced effects, and resource scaling in realistic quantum technologies. Other protocols involving more complex channel architectures, as, for instance, superpositions of Depolarizing channels can be investigated.   \\

\textbf{Acknowledgements}
We would like to thank the financial support from the Brazilian funding agencies Conselho Nacional de Desenvolvimento Cient\'{\i}fico e Tecnol\'ogico (CNPq), Funda\c{c}\~ao Carlos Chagas Filho de Amparo \`a Pesquisa do Estado do Rio de Janeiro (FAPERJ), Coordena\c{c}\~ao de Aperfei\c{c}oamento de Pessoal de N\'{\i}vel Superior (CAPES), National Institute for Science and Technology in Quantum Devices (INCT-QD / CNPq, Grant No. 408783/2024-9).





\bibliography{sample}


\end{document}